# Spatiotemporal Impact of Trade Policy Variables on Asian Manufacturing Hubs: Bayesian Global Vector Autoregression Model

Lutfu S. Sua
*Dept. of Management and Marketing, Southern University and A&M College, Baton Rouge, LA, USA*
Haibo Wang
*Dept. of International Business and Technology Studies, Texas A&M International University, Laredo, Texas, USA*
Jun Huang
*Dept. of Management and Marketing, Angelo State University, San Angelo, TX, USA*

**Abstract**

A novel spatiotemporal framework using diverse econometric approaches is proposed in this research to analyze relationships among eight economy-wide variables in varying market conditions. Employing Vector Autoregression (VAR) and Granger causality, we explore trade policy effects on emerging manufacturing hubs in China, India, Malaysia, Singapore, and Vietnam. A Bayesian Global Vector Autoregression (BGVAR) model also assesses interaction of cross unit and perform Unconditional and Conditional Forecasts. Utilizing time-series data from the Asian Development Bank, our study reveals multi-way cointegration and dynamic connectedness relationships among key economy-wide variables. This innovative framework enhances investment decisions and policymaking through a data-driven approach.

## 1. Introduction

Going through the challenges exacerbated by the COVID-19 pandemic, manufacturers worldwide learned the value of resiliency, flexibility, and efficiency. Managers needed to adapt faster and act with their supply chain partners to survive in a global business environment disrupted by shocks (Global Network of Advanced Manufacturing Hubs, 2022). Global manufacturing usually involves a long and complex process involving extensive supply chains to transform raw materials into final products supplied to end customers. Along these lines, global manufacturing networks that minimize costs and maximize efficiency are pervasively vulnerable to external shocks. In a world of frequent disruptions and shocks causing greater problems, managers and policymakers are considering new ways of improving the resiliency of global supply chains. The share of international trade based on labor-cost arbitrage is declining; global supply chains are leaning toward being more knowledge-intensive, relying more on high-skilled labor, and more regionally concentrated. Although the COVID-19 pandemic was the most recent and broadest interruption to manufacturing supply chains, several crises have exposed them and their national economies to shocks and disruptions (McKinsey Global Institute, 2020).

The United Nations Statistics Division reported that in 2018, China accounted for 30% of the global manufacturing output. However, the US-China trade war has forced many businesses to re-evaluate their global supply chains. Most multinational companies can move a quarter of their global product sources to other countries within the next 5 years. The COVID-19 pandemic, cyber-security concerns, and climate change risks are the main drivers of this trend. These uncertainties create opportunities for many countries to become the next manufacturing hub of the world (World Finance 2020). The COVID-19 crisis undoubtedly raised warning flags for policymakers regarding the potential of natural disasters inflicting economic damage on an unprecedented scale.

To examine the causality and cointegration effects of economy-wide policy variables on Asian manufacturing hubs, the following research questions are proposed:

- How do spillover effects affect economy-wide policy variables between Asian countries emerging as manufacturing hubs?
- What is the dynamic connectedness among manufacturing hubs in Asia?

This study aims to contribute to the current literature by developing a unique conceptual spatiotemporal framework to analyze causality, co-integration, and dynamic connectedness based on time-series data for China, India, Indonesia, Malaysia, and Vietnam, including key economy-wide policy variables. Spatiotemporal data analysis involves both space and time properties. Our model considers those five countries as spatial property and their time series as temporal property. Data availability and advances in computational technologies are the primary tools for undertaking spatiotemporal data analysis. Second, a standard Vector Auto Regression (VAR) estimation model is implemented for predicting potential manufacturing shifts and identifying the factors that would explain those shifts. Third, a Bayesian Global Vector Autoregression (BGVAR) model examines offshore decisions and identifies the dynamic connectedness among countries under diverse market conditions during these shifts.

The rest of this paper is structured as it follows. Section 2 provides a review of the trade policy and its impact on manufacturing hubs using causality, co-integration, and dynamic connectedness analyses. The econometric fundamentals are introduced in Section 3, followed by the results and findings in Section 4.

The implications for theory and practice are summarized in Section 5 along with the conclusions and future directions for research.

## 2. Literature Review

**2.1 Trade Policy**

Economic policy embodies a political ambition to lead the trade structure of a country toward economic growth and development. Although the impact of policy responses of the governments to the COVID-19 pandemic on financial markets is investigated to a certain extent (Sekandary and Bask, 2023; Deng et al., 2022; Roy, 2023), the question of how trade policies shape industrial output remains to be answered through sound research. The impact of economy-wide policy variables used in this study towards this aim is determined based on the following studies.

Analyzing the historical evolution of Indonesia's trade policy over the five decades, Puga and Venables (1999) study the trade policy's role in promoting industrialization and show that trade liberalization results in different industrial structures in countries with similar characteristics. Fanti and Buccella (2023) analyze which policy instrument should be selected along with the trade policy intervention results harmful for the national social welfare compared to the free trade regime while Grangnon (2022) investigates the role of financial development on non-resource tax revenue performance in developing economies by the means of international trade channel. Pangestu, Rahardja, and Ing (2015) argue that to improve competitiveness, its trade policy should shift from protected sectors towards promoting trade and industrial policies encouraging the movement of goods, services, and people. Sahakyan (2019) examines Canada's trade policy changes following multilateral liberalization and preferential trade agreements. Khalfaoui et al. (2022) utilize quantile vector autoregression to assess static and dynamic connectedness among green indexes across various market conditions. The effects of protectionist trade policies by the US on Canada in establishing trade relations with emerging markets such as China and India are also examined. Cascaldi-Garcia et al. (2020) measure risk, uncertainty, and volatility and their spillover effect on macroeconomic and financial outcomes around major events such as financial crises and pandemics. Mena et al. (2022) used a fuzzy-set qualitative comparative analysis to determine the factors leading to trade resilience amidst the COVID-19 pandemic. Wang et al. (2022) investigate the relationships between real expenditures on outbound tourism from China, international trade and economic growth, utilizing Bootstrap Autoregressive Distributed Lag approach.

Sections 2.2 and 2.3 elaborate on using cointegration, VAR, GVAR, and BGVAR models to assess the dynamic spillover relationship between the economy-wide policy variables.

## 2.2 Cross country (Unit) Interaction

A framework for implementing a volatility spillover metric based on forecast error variance decompositions using VARs was presented by Diebold and Yilmaz (2009). In order to expose connectedness system under both static and time-varying conditions, directional volatility spillovers were added to this framework, substantially improving it. To examine the economic relationships between the factors influencing trade policy, one of the econometric approaches used in this study is VAR for multi-way cointegration analysis.

Although several methodologies such as regression analysis, principal component analysis, co-skewness and co-kurtosis approaches, causality tests, and Markow models are used to demonstrate the spillover effects, Vector Auto Regression methods have become the mainstream for the analyses of spillover effects (Xu and Li, 2023). Numerous studies involved quantile connectedness in analyzing the spillover effect, particularly in stock markets (Xu and Li, 2023) and cryptocurrencies (Shahzad et al., 2023). Pham and Cepni (2022) use quantile connectedness to examine the spillovers between green bond performance and investor interest across extreme and normal market conditions. Gomez-Gonzalez et al. (2021) analyze the relationship between stock and oil market returns for seven countries and report substantial bidirectional spillover from stock to oil markets, while others have investigated the connectedness between volatility and economic policy uncertainty between metal market and oil shocks (Oliyide et al. 2021) and the dynamic spillovers between South African and international equity markets (Fowowe and Shuaibu, 2016).

Examining dynamic network spillovers and the impacts of major events on relevant networks have recently received increasing attention, resulting in a number of versions of VAR models (Diebold and Yilmaz, 2012). BGVAR is a Bayesian extension of the global vector autoregressive (GVAR) model which is a coherent multi-country framework that makes it possible to account for the second- and higher-order spillover effects– and spillovers through third-countries (Feldkircher and Huber, 2016). GVAR a global modelling framework considered as an extension of VAR for analyzing the international macroeconomic transmission of shocks while accounting for drivers of economic activity, interlinkages and spillovers between different countries, and the effects of unobserved or observed common factors (Mohaddes and Raissi, 2020). Eickmeier and Ng (2015) utilize GVAR to model financial variables jointly with macroeconomic variables in 33 countries over a 27-year period. By pooling the forecasts obtained from different GVAR models estimated over alternative estimation periods, Pesaran et al. (2009) analyze the effects of model and estimation uncertainty on forecast outcomes.

Tiwari et al. (2022) use a rolling window-based QVAR when looking for the conditional volatility spillover before and during COVID-19. Antonakakis et al. (2019) extended this model to investigate uncertainty spillovers between Greece and Europe through the global fiscal crisis. Building upon the work of (Diebold & Yilmaz, 2012); Gabauer et al. (2023) propose a model free of unconditional connectedness measures, investigating only the bivariate relations, thus significantly increasing the network size. Florian (2016) develops BGVAR with stochastic volatility in order to improve the existing homoscedastic framework forecasting a set of macroeconomic variables. Dovern et al. (2016) decompose the predictive joint density into its marginals and a copula term identifying the dependence relationship across countries to study the source of performance gains, using BGVAR. Gupta et al. (2020) analyzed co-existing and temporal causal relationships of uncertainty at the state level by utilizing BGVAR while Cuaresma et al. (2016) developed BGVAR models to forecast a set of macroeconomic and financial variables, reporting superior results over country specific vector autoregressions.

The review of the literature indicates that while there is indeed research that focuses on only one country or uses simple regression models with a few variables, few works are reporting a system-wide experiment on the region using BGVAR to analyze time-series data in terms of cointegration and dynamic connectedness. This research aims to fill this gap by analyzing the impact of several economy-wide policy variables on Asian manufacturing hubs by employing a spatiotemporal analysis.

## 3. Methodology

The workflow of time-series analysis is illustrated in Figure 1.

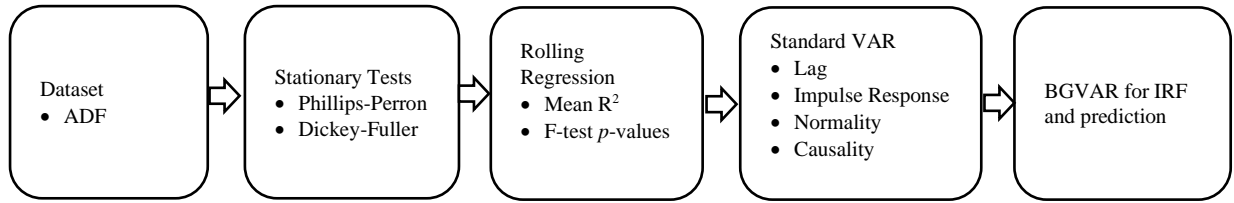

Figure 1. Workflow for time-series data analysis

Note: A five-step flowchart showing the analysis steps and the test used in each step.

Phillips-Perron and Dickey-Fuller unit root tests are employed to confirm the stationarity of the time series (Cheung and Lai, 1995 and Kipiński, 2007). The null hypothesis posits that the time series is stationary. However, if the p-value is lower than the designated significance level, it suggests that the series is non-stationary. To confirm stationarity, both tests are applied. Interestingly, occasional discrepancies in the test results indicate that the series may be trend-stationary rather than strictly stationary.

Rolling regression analyses are utilized to examine relationships among variables. They are one of the simplest ways to analyze changing relationships over time. Rolling regression uses linear regression but allows the data set to change over time. Rolling regression estimates model parameters using a fixed window of time over the entire data set (Zanin & Marra, 2012). They are implemented by the Rolling Ordinary Least-Squares (OLS) function.

The standard OLS formula is given as:

$$Min\ S = \sum_{i=1}^{n}(\hat{\epsilon}_i)^2 = \sum_{i=1}^{n}(y_i - \hat{y}_i)^2 = \sum_{i=1}^{n}(y_i - b_i x_i - b_0)^2 \tag{1}$$

Where, $\hat{y}_i$ is the $i$th observation's predicted value, $y_i$ is the $i$th observation's actual value, $\hat{\epsilon}_i$ is the $i$th observation's error/residual, and $n$ is the number of observations. Then the partial derivative for each coefficient is set to zero to compute the values of $b_i$ and $b_0$. There are several assumptions regarding the data properties involved in testing rolling regression on the time series data. First, the dependent and independent variables need to be stationary, and the statistical properties of the sample under each time interval would reveal the average statistical properties of the data within the time interval. This assumption is satisfying since we slide the window across the dataset instead of sampling from the whole dataset and taking the log on the data used for calculation. If the dependent and independent variables are co-integrated, then the results of a rolling regression are valid even if the data are not stationary. The second assumption is that the mean value of the residuals of the estimated rolling OLS model is equal to zero, and there is no perfect multicollinearity. This assumption is applied to any regression model.

The VAR model identifies the relationship between the variables over a period and it is expressed as below:

$$x_t = \alpha + \emptyset_1 x_{t-1} + \ldots + \emptyset_n x_{t-n} + w_t \tag{2}$$

$$w_t \sim Normal(0, \sum_w)$$

Causality and cointegration are assessed before and after each major event. Johansen test is used here to test the cointegration relationship (Johansen, 1991).

The Granger test assesses causality among the variables being examined. The null hypothesis states that one variable does not explain the variation in another. Probabilities lower than a pre-specified level spark the rejection of the null hypothesis, resulting in causality between the pair of variables. Granger causality model not only calculates the predictive precision of the variable of interest based on its past values but also on the moving average of past values of the second variable. To assess the Granger-causality model's goodness of fit, the root mean square (RMS) is computed assuming that the time series' residuals are randomly distributed and normally distributed.

$$RMS = \sqrt{\frac{\sum_{i=1}^{N-L}(\check{x}_i - x'_i)^2}{N-L}} \tag{3}$$

where $L$ is the model's order, $N$ is the number of data points in the time series, and $N-L$ is the number of data points used to build the model. $\check{x}$ is the vector for the data points subtracted from the original data by the first $L$ data.

Jarque-Bera (JB) goodness-of-fit test evaluates the cointegration between multiple variables to assess the distribution of the VAR model's residuals for the variables. Skewness and kurtosis values are used to check if the residuals are randomly distributed for normality. $p$-values $< 0.05$ shows the VAR model's validity, thus a bidirectional cointegration. Pairwise cointegration between the variables implies the existence of system-wide cointegration.

$$\begin{bmatrix} cpi_t \\ emp_t \\ gdp_t \\ mpi_t \\ exp_t \\ imp_t \\ exc_t \\ mny_t \end{bmatrix} = A_0 + A_1 \begin{bmatrix} cpi_{t-1} \\ emp_{t-1} \\ gdp_{t-1} \\ mpi_{t-1} \\ exp_{t-1} \\ imp_{t-1} \\ exc_{t-1} \\ mny_{t-1} \end{bmatrix} + A_2 \begin{bmatrix} cpi_{t-2} \\ emp_{t-2} \\ gdp_{t-2} \\ mpi_{t-2} \\ exp_{t-2} \\ imp_{t-2} \\ exc_{t-2} \\ mny_{t-2} \end{bmatrix} + \cdots + A_p \begin{bmatrix} cpi_{t-p} \\ emp_{t-p} \\ gdp_t \\ mpi_{t-p} \\ exp_{t-p} \\ imp_{t-p} \\ exc_{t-p} \\ mny_{t-p} \end{bmatrix} + CD_t + u_t \qquad (4)$$

*where:*

- $cpi_t$: Consumer Price Index (CPI)
- $emp_t$: Unemployment
- $gdp_t$: GDP per capita
- $mpi_t$: Manufacturing Price Index (MPI)
- $exp_t$: Annual Change in Exports
- $imp_t$: Annual Change in Imports
- $exc$: Exchange rate
- $mny_t$: Money supply

The GVAR comprises two layers via which the model can identify cross-country spillovers. Separate time series models are estimated in the first layer for each country included in the global model. This accounts for the cross-country differences of the economies since homogeneity is not imposed. The country models are stacked in the second layer resulting in a global model capable of tracing the spatial propagation of a shock as well as its time dynamics (Feldkircher and Huber, 2016). The key to the GVAR is the systematic inclusion of the country-specific foreign variables into the individual country models as a way of dealing with the common factor dependencies that exist in the global economy (Dees et al., 2007).

The VARX*(1,1) model for country $i \in 0, \ldots, N$ is given by

$$x_{i,t} = a_{i0} + a_{i1}t + \psi_{i1}x_{i,t-1} + \Lambda_{i0}x^*_{i,t} + \Lambda_{i1}x^*_{i,t-1} + \varepsilon_{i,t} \qquad (5)$$

where $x^*_{i,t}$ is a $k_i x 1$ endogenous variables matrix in country $i$ at time $t$, $a_{i0}$ denotes the coefficient on the constant and $a_{i1}$ is the coefficient on the deterministic time trend. The $k_i \times k_i$ matrix of dynamic coefficients for the lagged endogenous variables in country $i$ is given by $\psi_{i1}$. The right-hand side of (5) features weakly exogenous which can be defined as

$$x_{i,t}^* = \sum_{j \neq i}^{N} \omega_{ij} x_{j,t}, \tag{6}$$

where $x_{i,t}^*$ is of dimension $k_i^* \times 1$ and $\omega_{ij}$ denotes bilateral weights between countries $i$ and $j$. These are most often based on trade or financial flows in empirical applications. The corresponding $k_i \times k_i^*$-dimensional parameter matrices are provided by $\Lambda_{il}(l = 0,1)$. The usual vector white noise process is denoted by $\varepsilon_{i,t} \sim N(0, \Sigma_{\varepsilon,i})$, with variance–covariance matrix $\Sigma_{\varepsilon,j}$.

It should be noted that concurrently entering the model are weakly exogenous variables. Since bilateral weights $\omega_{ij}$ are assumed to be fixed and exogenous, weakly exogenous variables resemble a function of $x_t$ and are thus endogenously determined within the global system. More of truly exogenous variables can be added in a straightforward fashion. Second, note that if $\Lambda_{i0} = \Lambda_{i1} = 0$ the VARX* collapses to a standard first-order VAR featuring a deterministic time trend.

Solving the country-specific VARX models simultaneously for all domestic variables starts by stacking x $x_{i,t}$ and $x_{i,t}^*$ to retrieve a $(k_i + k_i^*)$ −dimensional vector $z_{i,t} = (x_{i,t}', x_{i,t}^{*'})'$.

Collecting all contemporaneous terms on the left-hand side, the model can be reformulated in (5) as follows:

$$A_i z_{i,t} = a_{i0} + a_{i1} t + B_i z_{i,t-1} + \varepsilon_{i,t}, \tag{7}$$

with $A_i = (I_{k_i} - \Lambda_{i0})$ and $B_i = (\Psi_{i1}, \Lambda_{i1})$ being $k_i \times (k_i + k_i^*)$ matrices. Note that by utilizing a suitable $(k_i + k_i^*) \times k$ link matrix $W_i$, where $k = \sum_{i=0}^{N} k_i$ denotes the number of endogenous variables in the global system, $z_{i,t}$ can be rewritten in terms of a global vector. The $k$-dimensional global vector $x_t = ((x_{0,t}', \ldots, (x_{N,t}')'$ contains all endogenous variables in the system. Thus, it is easy to show that by utilizing (6), $z_{i,t}$ can also be written as

$$z_{i,t} = W_i x_t \tag{8}$$

This allows re-expression of the country model in (5) in terms of the global vector,

$$A_i W_i x_t = a_{i0} + a_{i1} t + B_i W_i x_{t-1} + \varepsilon_{i,t} \tag{9}$$

Stacking the $A_i W_i$ and $B_i W_i$ matrices for all countries leads to

$$G x_t = a_0 + a_1 t + H x_{t-1} + \varepsilon_t \tag{10}$$

where

$a_0 = (a_{00}', \ldots, a_{N0}')'$, $a_1 = (a_{01}', \ldots, a_{N1}')'$, $G = [(A_0 W_0)', \ldots, (A_N W_N)']$, $H = [(B_0 W_0)', \ldots, (B_N W_N)']$ and $\varepsilon_t = (\varepsilon_{0,t}', \ldots, \varepsilon_{N,t}')' \sim N(0, \Sigma_\varepsilon)$. It is assumed that $\Sigma_\varepsilon$ is a block-diagonal matrix with $\Sigma_{\varepsilon,i}$ on the main diagonal. Multiplying from the left by $G^{-1}$ results in the global vector autoregressive model:

$$x_t = G^{-1}a_0 + G^{-1}a_1 t + G^{-1}Hx_{t-1} + G^{-1}\varepsilon_t = b_0 + b_1 t + Fx_{t-1} + e_t, \quad (11)$$

with F denoting the $k \times k$ companion matrix and $e_t \sim N(0, \Sigma_\varepsilon)$ where $\Sigma_\varepsilon = G^{-1}\Sigma_\varepsilon G^{-1}$. Thus, it can easily be observed that the matrix G establishes contemporaneous relationships between countries. The Bayesian treatment GVAR allows inclusion of large information sets by mitigating overfitting-related issues which improves inference leading to better out-of-sample forecasts (Böck et al., 2020).

BGVAR is implemented in this research to measure the dynamic connectedness among different variables by evaluating the average impact of a shock in each variable has on all the other variables under diverse market conditions.

## 4. Empirical findings

### 4.1 Data

Time series data are obtained from a variety of sources for the causality, cointegration, and connectedness analyses. CPI, Unemployment Rate (Emp.), GDP per capita (GDP), MPI, Exports (Exp), Imports (Imp), Exchange Rate (Exc), and Money Supply (Mny) data are obtained from the Asian Development Bank.

The causality, cointegration, and dynamic connectedness analyses are implemented in R language. Table 1 presents the non-stationary factors for each country.

Table 1. Results of stationarity tests on variables

|  | CPI | Emp. | GDP | MPI | Exp. | Imp. | Exc. | Mny. |
|---|---|---|---|---|---|---|---|---|
| China | 0.04** | 0.01*** | 0.80 | 0.07 | 0.09 | 0.04 | 0.88 | 0.12 |
| India | 0.65 | 0.57 | 0.86 | 0.22 | 0.02 | 0.15 | 0.78 | 0.47 |
| Malaysia | 0.44 | 0.90 | 0.01*** | 0.57 | 0.01*** | 0.01*** | 0.80 | 0.16 |
| Singapore | 0.77 | 0.15 | 0.01*** | 0.29 | 0.01*** | 0.01*** | 0.92 | 0.25 |
| Vietnam | 0.26 | 0.58 | 0.99 | 0.51 | 0.01*** | 0.01*** | 0.69 | 0.01*** |

Stationarity; *p-value<=0.1, ** p-value<=0.05, *** p-value<=0.01

### 4.2 Causality, Cointegration, and Regression Results

A standard VAR model is assessed based on the variables, and the results are provided in Table 2 and Table 3. The findings in VAR suggest significant relationships among the factors.

Table 2. Granger Causality results

| Country |  | CPI | Emp. | GDP | MPI | Exp | Imp. | Exc. | Mny. |
|---|---|---|---|---|---|---|---|---|---|
| China | CPI | - | - | - | - | 0.005578*** | 0.001325*** | - | - |
|  | Emp. | - | - | - | 0.052460* | - | - | - | - |
|  | GDP | - | - | - | - | 0.082840* | 0.057170* | - | - |
|  | MPI | - | - | - | - | - | - | - | - |
|  | Exp. | 0.03034** | - | - | - | - | - | - | - |
|  | Imp. | 0.03178** | - | - | - | - | - | - | - |
|  | Exc. | - | - | - | - | 0.008668*** | 0.027640** | - | - |
|  | Mny. | - | - | - | - | 0.016170** | 0.005055*** | - | - |
| India | CPI | - | - | - | - | - | - | - | - |
|  | Emp. | - | - | 4.33e-05*** | - | 0.029420** | 0.013770** | - | - |

| Country | | CPI | Emp. | GDP | MPI | Exp | Imp. | Exc. | Mny. |
|---|---|---|---|---|---|---|---|---|---|
| | GDP | - | 0.048780** | - | - | 0.046100** | 0.034020** | - | - |
| | MPI | - | - | - | - | - | - | - | 0.008566*** |
| | Exp. | - | - | - | - | - | - | - | - |
| | Imp. | - | - | - | - | - | - | - | - |
| | Exc. | - | - | - | 0.058300* | - | - | - | - |
| | Mny. | 0.000320*** | - | - | - | - | - | - | - |
| Malaysia | CPI | - | - | 0.010680** | - | 0.003826*** | - | - | - |
| | Emp. | - | - | - | - | 0.000505*** | 0.006071*** | 0.077420* | - |
| | GDP | - | - | - | - | 0.002757*** | - | - | - |
| | MPI | - | - | - | - | - | - | - | - |
| | Exp. | - | - | - | - | - | - | - | 0.024670** |
| | Imp. | - | - | - | - | 0.001727*** | - | - | - |
| | Exc. | - | 0.047560** | - | - | - | - | - | - |
| | Mny. | - | - | - | - | - | - | - | - |
| Singapore | CPI | - | - | - | - | 0.004437*** | 0.019020** | 0.030130** | 0.045940** |
| | Emp. | - | - | - | 0.086790* | 0.009131*** | 0.043720** | - | - |
| | GDP | - | - | - | 0.096010* | - | - | - | - |
| | MPI | 0.053260* | 0.024040** | 0.013670** | - | - | 0.001721*** | 0.092370* | - |
| | Exp. | - | - | - | - | - | 0.002336*** | - | - |
| | Imp. | - | - | - | 0.015960** | 0.001301*** | - | - | - |
| | Exc. | 0.033880** | - | - | - | 0.01369** | 0.07226* | - | - |
| | Mny. | - | - | - | - | - | - | - | - |
| Vietnam | CPI | - | - | - | - | 0.001207*** | 0.001204*** | - | - |
| | Emp. | - | - | - | - | - | - | - | - |
| | GDP | 0.052420* | - | - | 0.056760* | 0.023110** | - | - | - |
| | MPI | - | - | 0.006825*** | - | - | - | - | - |
| | Exp. | 0.070830* | - | - | - | - | 0.085380* | 0.09479* | - |
| | Imp. | 0.012420** | - | - | - | - | - | 0.08476* | - |
| | Exc. | - | - | - | - | - | - | - | - |
| | Mny. | 0.044100** | - | - | - | 0.009139*** | 0.022110** | - | - |

*Note: \*\*\* at 1%, \*\*at 5%, and\*at 10% significance levels.*

The Johansen test is utilized to evaluate the validity of cointegration relationships, as presented in Table 3. Having only two univariate time series, we can only have two ranks: $r \leq 1$ or $r = 0$, which shows whether cointegration exists ($r = 1$) or not ($r \leq 0$).

The hypothesis is stated as:

$H_0$: no cointegration

$H_1$: $H_0$ does not hold

The first hypothesis, $r = 0$, tests for cointegration. Thus, a footnote as Trace statistics of $r = 0$ is reported in Table 3 (\*\*\* at 1%, \*\* at 5%, and \* at 10% levels of significance).

Table 3. Cointegration results

| **Country** | | CPI | Emp. | GDP | MPI | Exp | Imp. | Exc. | Mny. |
|---|---|---|---|---|---|---|---|---|---|
| China | CPI | - | 40.94*** | 37.75*** | 41.66*** | 39.91*** | 42.79*** | 31.59*** | 36.28*** |
| | Emp. | - | - | - | 18.28* | 31.53*** | 35.31*** | 18.37* | 21.47** |
| | GDP | - | - | - | - | 17.86* | 20.67** | - | - |
| | MPI | - | - | - | - | 32.26*** | 34.31*** | - | 22.47** |
| | Exp. | - | - | - | - | - | 37.87*** | 21.15** | 30.47*** |
| | Imp. | - | - | - | - | - | - | 23.53** | 37.87*** |
| | Exc. | - | - | - | - | - | - | - | - |
| | Mny. | - | - | - | - | - | - | - | - |
| India | CPI | - | 19.67* | - | 34.47*** | 23.90** | 23.87** | 21.91** | - |
| | Emp. | - | - | 32.69*** | 21.31** | 28.32*** | 28.13*** | 20.99** | 20.27** |
| | GDP | - | - | - | - | 19.91* | 20.93** | 22.40** | - |
| | MPI | - | - | - | - | 27.29*** | 24.49** | 18.79* | 26.43*** |
| | Exp. | - | - | - | - | - | 39.14*** | 22.76** | 23.60** |
| | Imp. | - | - | - | - | - | - | 19.23* | 25.56*** |
| | Exc. | - | - | - | - | - | - | - | 23.53** |

|  |  |  |  |  |  |  |  |  |
|---|---|---|---|---|---|---|---|---|
|  | Mny. | - | - | - | - | - | - | - | - |
| Malaysia | CPI | - | 19.07* | 22.15** | 18.55* | 19.88* | 21.65** | - | 22.79** |
|  | Emp. | - | - | 20.83** | - | 36.98*** | 33.91*** | - | 26.76*** |
|  | GDP | - | - | - | 23.20** | 40.55*** | 38.48*** | 21.64** | 31.77*** |
|  | MPI | - | - | - | - | 27.75*** | 25.12*** | 18.02* | 31.36*** |
|  | Exp. | - | - | - | - | - | 35.16*** | 24.12** | 32.27*** |
|  | Imp. | - | - | - | - | - | - | 22.25** | 33.61*** |
|  | Exc. | - | - | - | - | - | - | - | 24.96*** |
|  | Mny. | - | - | - | - | - | - | - | - |
| Singapore | CPI | - | 18.42* | 21.77 | - | 21.59** | - | - | 24.81*** |
|  | Emp. | - | - | 32.16*** | 28.99*** | 45.58*** | 35.44*** | 20.49** | 25.50*** |
|  | GDP | - | - | - | 29.91*** | 55.79*** | 47.69*** | 24.33** | 36.09*** |
|  | MPI | - | - | - | - | 42.91*** | 37.15*** | - | 26.63*** |
|  | Exp. | - | - | - | - | - | 36.84*** | 21.51** | 29.30*** |
|  | Imp. | - | - | - | - | - | - | 23.00** | 28.94*** |
|  | Exc. | - | - | - | - | - | - | - | 24.37** |
|  | Mny. | - | - | - | - | - | - | - | - |
| Vietnam | CPI | - | 28.71*** | 31.17*** | 27.20*** | 28.85*** | 34.28*** | 29.89*** | 39.40*** |
|  | Emp. | - | - | - | - | 28.38*** | 36.01*** | 23.36** | 46.68*** |
|  | GDP | - | - | - | - | - | 21.47** | - | 35.60*** |
|  | MPI | - | - | - | - | 24.73*** | 31.00*** | - | 29.52*** |
|  | Exp. | - | - | - | - | - | 32.15*** | 31.36*** | 38.74*** |
|  | Imp. | - | - | - | - | - | - | 37.88*** | 47.62*** |
|  | Exc. | - | - | - | - | - | - | - | 37.01*** |
|  | Mny. | - | - | - | - | - | - | - | - |

*Note: \*\*\* at 1%, \*\*at 5%, and \*at 10% significance levels.*

Tables 2 and 3 summarize the causality and co-integration relations between the variables. There is compelling evidence to reject the null hypothesis of no cointegration between the variables, with a few exceptions. There is no cointegration between Money Supply and GDP per capita in India and China. Another interesting finding is that the Exchange Rate has no cointegration relationship with MPI and GDP per capita in China or Vietnam. Finally, the results indicate that unemployment and MPI do not cointegrate in Malaysia and Vietnam.

Rolling regression results among the factors for each country are reported in Table 4.

Table 4. Rolling regression results ($R^2$ values)

|  |  | CPI | Emp. | GDP | MPI | Exp | Imp. | Exc. | Mny. |
|---|---|---|---|---|---|---|---|---|---|
| China | CPI | - | -0.10870 | 0.04727 | 0.11900 | 0.34330 | 0.39720 | 0.102 | 0.04834 |
|  | Emp. |  | - | 0.09258 | 0.11490 | 0.12350 | 0.09736 | 0.05718 | 0.07018 |
|  | GDP |  |  | - | -0.09518 | **0.16230\*** | **0.15560\*** | 0.06892 | 0.01739 |
|  | MPI |  |  |  | - | 0.10010 | 0.07858 | 0.05906 | 0.10600 |
|  | Exp. |  |  |  |  | - | 0.04429 | 0.1298 | 0.03735 |
|  | Imp. |  |  |  |  |  | - | 0.1321 | 0.04546 |
|  | Exc. |  |  |  |  |  |  | - | 0.16520 |
|  | Mny. |  |  |  |  |  |  |  | - |
| India | CPI | - | -0.08997 | -0.10600 | -0.02756 | **0.17810** | 0.05040 | -0.00399 | -0.07941 |
|  | Emp. |  | - | **0.67100\*\*** | -0.07511 | **0.33860\*** | **0.26660\*** | -0.1961 | 0.02428 |
|  | GDP |  |  | - | -0.07938 | **0.31110\*** | 0.20400 | 0.009567 | 0.03986 |
|  | MPI |  |  |  | - | 0.15570 | -0.01030 | -0.0697 | **0.33070\*** |
|  | Exp. |  |  |  |  | - | -0.00450 | -0.05455 | 0.02408 |
|  | Imp. |  |  |  |  |  | - | -0.1038 | 0.02923 |
|  | Exc. |  |  |  |  |  |  | - | 0.04889 |
|  | Mny. |  |  |  |  |  |  |  | - |
| Malaysia | CPI | - | -0.11710 | 0.04210 | 0.01499 | **0.52010\*\*** | **0.52950\*\*** | 0.02747 | 0.08546 |
|  | Emp. |  | - | 0.05816 | -0.06794 | **0.48840\*\*** | **0.54450\*\*** | 0.02354 | 0.09773 |
|  | GDP |  |  | - | 0.06042 | **0.74780\*\*** | **0.70940\*\*** | 0.07144 | 0.07731 |
|  | MPI |  |  |  | - | **0.31030\*** | **0.37970\*\*** | -0.01936 | 0.07565 |
|  | Exp. |  |  |  |  | - | **0.24070\*\*** | -0.02279 | **0.30230\*\*** |
|         | Imp. |   |          |          |          |           | -         | -0.02317  | **0.19210** |
|         | Exc. |   |          |          |          |           | -         |           | 0.09766   |
|         | Mny. |   |          |          |          |           |           |           | -         |
| Singapore | CPI | - | 0.05952 | 0.18750 | 0.13280 | **0.51600**\*\* | **0.43130**\*\* | **0.22160**\*\* | **0.26480**\*\* |
|         | Emp. |   | -        | **0.26130**\* | 0.05811 | **0.34870**\*\* | **0.29460**\*\* | -0.008378 | 0.14190   |
|         | GDP  |   |          | -        | 0.04963 | 0.14730   | 0.12290   | 0.03620   | 0.09044   |
|         | MPI  |   |          |          | -        | 0.05626   | **0.56270**\*\* | 0.14160   | 0.07964   |
|         | Exp. |   |          |          |          | -         | **0.45780**\*\* | -0.00200  | 0.08992   |
|         | Imp. |   |          |          |          |           | -         | -0.00940  | 0.07992   |
|         | Exc. |   |          |          |          |           |           | -         | 0.1069    |
|         | Mny. |   |          |          |          |           |           |           | -         |
| Vietnam | CPI  | - | -0.07163 | -0.10770 | -0.05453 | **0.45330**\*\* | **0.43630**\*\* | **0.25490**\*\* | -0.02499  |
|         | Emp. |   | -        | -0.10530 | -0.10450 | 0.05280   | 0.05636   | **0.20340** | 0.03006   |
|         | GDP  |   |          | -        | 0.09047  | **0.28870**\*\* | **0.44190**\*\* | 0.23970   | -0.04507  |
|         | MPI  |   |          |          | -        | 0.05334   | 0.06749   | 0.24700   | -0.04447  |
|         | Exp. |   |          |          |          | -         | **0.17460** | **0.42600**\* | 0.01816   |
|         | Imp. |   |          |          |          |           | -         | **0.43380**\* | **0.36030**\* |
|         | Exc. |   |          |          |          |           |           | -         | 0.00205   |
|         | Mny. |   |          |          |          |           |           |           | -         |

Table 4 reports the $R^2$ values of the rolling regression analysis. Numbers in bold have F-test $p$-values <=0.1, with * means $p$-values <=0.05, with ** means $p$-values <=0.01. Since the rolling regression model is sensitive to outliers and the prediction powers are affected by the rolling window size, the mean $R^2$ values tend to be low on the small rolling window size.

### 4.3 BGVAR Results

The BGVAR model uses the NG prior (Normal-Gamma prior) which is a type of Bayesian shrinkage prior often used to estimate the parameters in VAR models, particularly when dealing with large systems where regularization is needed to prevent overfitting. Table 5 provides a summary of BGVAR model's outputs which offer insights into the model diagnostics, convergence, correlations, and residuals.

Table 5. Model Summary

| Model Information | | |
|---|---|---|
| Prior: Normal-Gamma prior (NG) | | |
| Number of lags for endogenous variables: 1 | | |
| Number of lags for weakly exogenous variables: 1 | | |
| Number of posterior draws: 1000/1=1000 | | |
| Number of stable posterior draws: 100 | | |
| Number of cross-sectional units: 5 | | |
| Convergence diagnostics | | |
| Geweke statistic: | | |
| 280 out of 1680 variables' z-values exceed the 1.96 threshold (16.67%). | | |
| F-test, first order serial autocorrelation of cross-unit residuals | | |
| Summary statistics | | |
|  | p-values | % |
| > 0.10 | 31 | 77.5 |
| 0.05 – 0.10 | 1 | 2.5 |
| 0.01 – 0.05 | 3 | 7.5 |
| < 0.01 | 5 | 12.5 |
| Average pairwise cross-unit correlation of unit-model residuals | | |
| Summary statistics | | |

|         | CPI      | Unemp. | GDP | MPI | Exp. | Imp. | Exchg. | Mny. |
|---------|----------|--------|-----|-----|------|------|--------|------|
| < 0.1   | 4 (80%)  | 4      | 1   | 3   | 2    | 3    | 1      | 1    |
| 0.1 – 0.2 | 0 (0%) | 0      | 1   | 2   | 2    | 1    | 2      | 3    |
| 0.2 – 0.5 | 1 (20%)| 1      | 3   | 0   | 1    | 1    | 2      | 1    |
| > 0.5   | 0 (0%)   | 0      | 0   | 0   | 0    | 0    | 0      | 0    |

It can be observed that 77.5% of p-values are >0.1, indicating no significant autocorrelation in the majority of cross-unit residuals. While most residuals are well-behaved, a small subset shows autocorrelations which can be addressed by introducing additional explanatory variables. Majority of the residual correlations across variables are less than 0.1, suggesting the model adequately accounts for interdependencies among the countries. Higher correlations are observed for exchange rate and GDP per capita, indicating stronger interconnections among these countries, likely due to trade and investment relations. Weak correlations suggest that the model effectively captures country-specific shocks.

MPI, CPI, unemployment, exports, and imports show predominantly weak residual correlations, suggesting well-isolated dynamics for these variables. Meanwhile, higher correlations for money supply and exchange indicate shared financial influences in the region. In light of these results, the model appears robust for most variables, with weak cross-unit correlations suggesting that country-specific shocks are well-captures. The majority of residuals lack significant autocorrelations, supporting the validity of short-term dynamics.

The normalized trade matrix in Table 6 shows the trade intensities between the countries. The values indicate the normalized trade intensity, normalized such that higher values reflect stronger relationships.

Table 6. Normalized trade matrix

| From/To   | China  | Indonesia | Malaysia | Singapore | Vietnam |
|-----------|--------|-----------|----------|-----------|---------|
| China     | 0      | 0.7173    | 0.9806   | 0.6891    | 1.0000  |
| Indonesia | 0.6844 | 0         | 0.0726   | 0.1330    | 0.0219  |
| Malaysia  | 0.5861 | 0.0619    | 0        | 0.4591    | 0.0351  |
| Singapore | 0.8220 | 0.1213    | 0.7271   | 0         | 0.0737  |
| Vietnam   | 0.7179 | 0         | 0.0198   | 0.0101    | 0       |

It should be noted that the diagonal values are zero since intra-country trade is excluded in this study. It can be observed that China is a dominant trade partner for all other countries, Vietnam showing the strongest dependency (1.0000), followed by Malaysia and Indonesia. Meanwhile, Singapore shows a strong relationship with Malaysia and China, reflecting its role as a regional trade hub. Also, trade from Indonesia and Vietnam to other countries is relatively low, except for their relationships with China. The high values for China indicate its centrality in the trade network. Policies or shocks affecting its trade will have substantial highlights. Singapore's high connectivity with China and Malaysia highlights its role as a key intermediary in the region. Vietnam's strong trade dependency on China suggests a significant exposure to China's economic conditions.

Figure 2 represents the unconditional forecast (n.ahead=5) of the MPI for the five Southeast Asian countries.

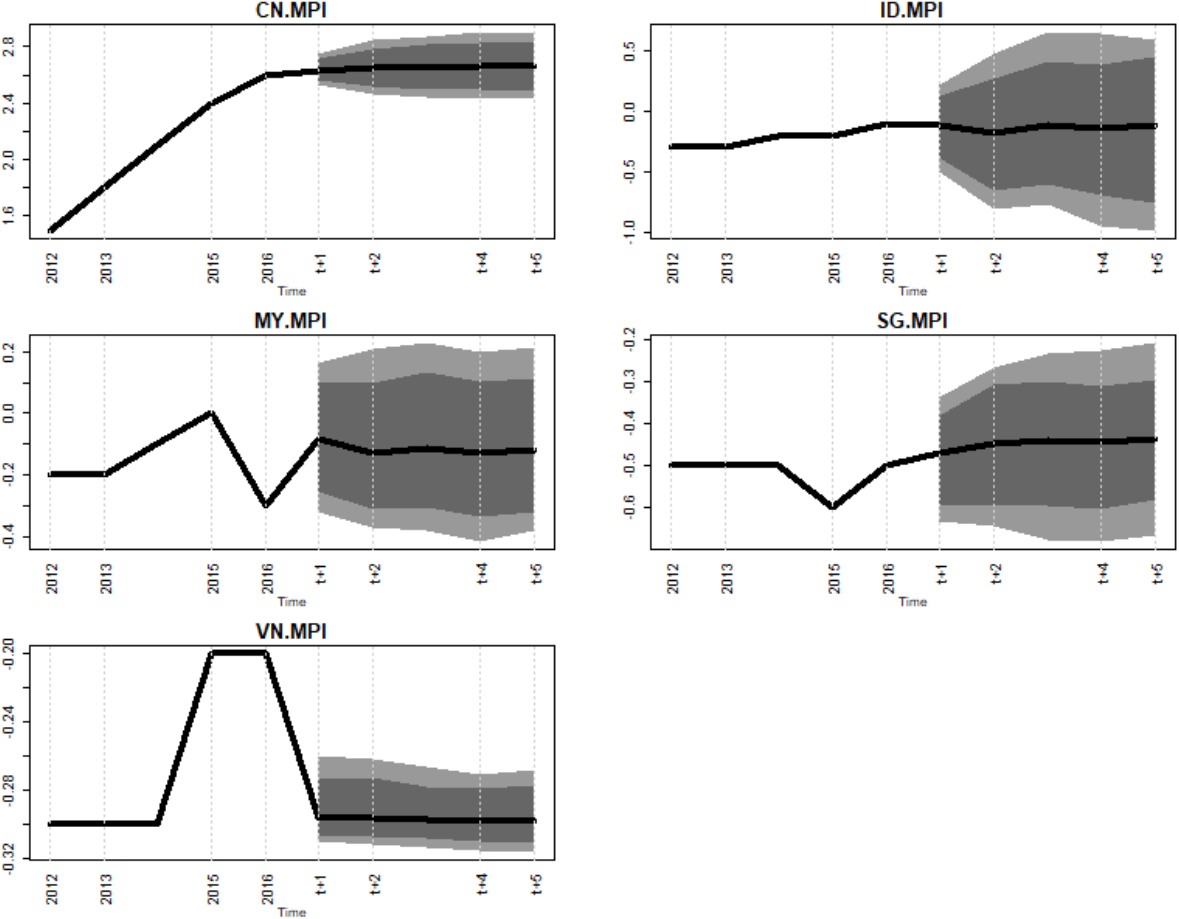

Figure 2. Unconditional forecast

This forecast is based on the BGVAR model's historical data and parameters, without imposing constraints or assumptions on future values. The shaded areas around the forecast lines represent uncertainty (confidence intervals), with darker shades indicating higher certainty. All countries exhibit widening confidence intervals over the 5 forecast periods, which is typical in BGVAR predictions as uncertainty grows with time. China's MPI forecast stands out due to its clear upward trajectory and low uncertainty, stabilizing slightly towards the end of the forecast period while others are more variable or stable.

The model forecasts minimal uncertainty, reflecting high confidence in the upward trajectory. A moderate increase is observed for Indonesia after a small dip, with growing uncertainty as the forecast progresses.

For Malaysia, MPI fluctuates in the past, and the forecast suggests stabilization with some uncertainty. A steady improvement is observed for Singapore, but higher uncertainty in the forecast while MPI stabilizes after a sharp increase in earlier periods in Vietnam, with moderate uncertainty.

The conditional forecasts (using fixed and uncertain constraints for China's MPI) provide insights into how imposing specific assumptions on China's MPI impacts the MPI of other countries.

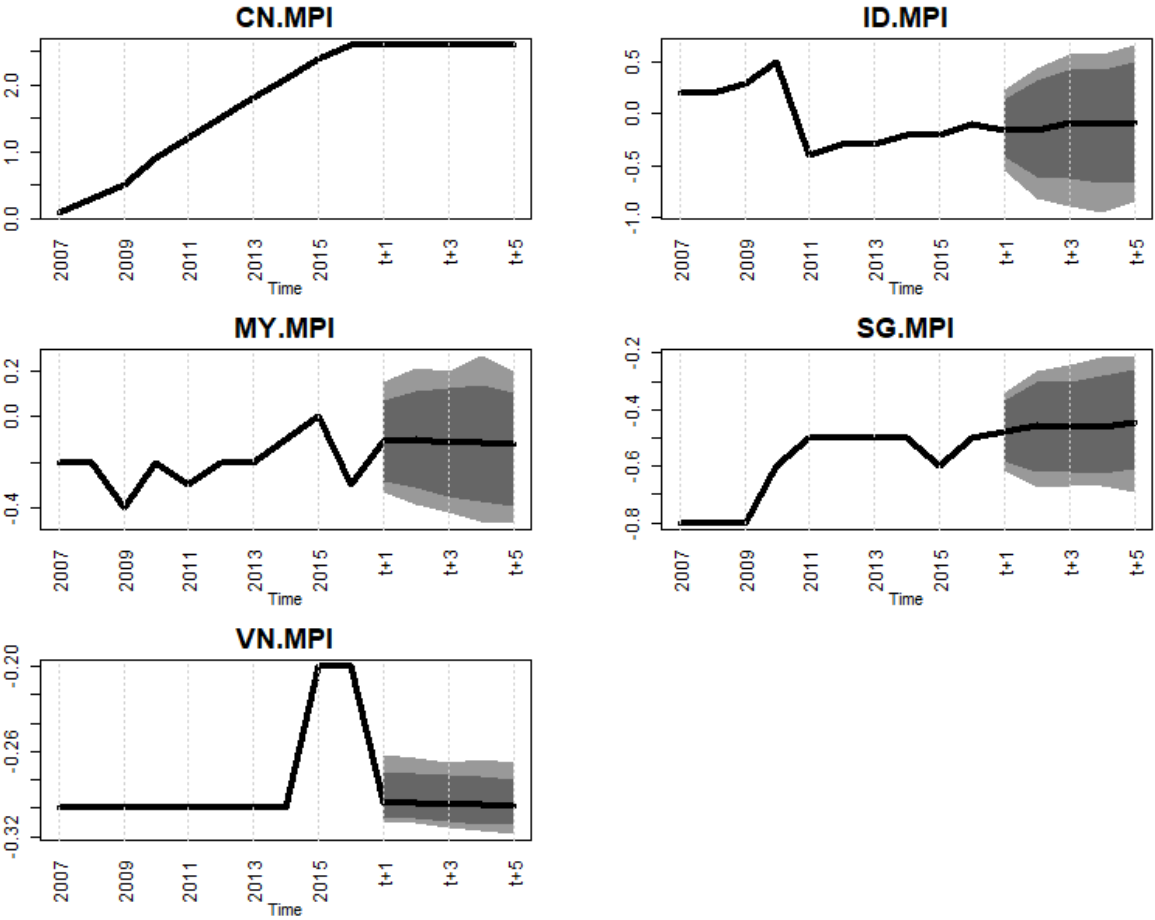

Figure 3. Conditional forecast with fixed constraints

This allows a small uncertainty around the future values of China's MPI. This uncertainty range is set to 0.1%. The forecasts of other countries' MPIs are based on this condition. The future values of China's MPI are fixed at the last observed value for the next 5 periods. This scenario assumes no further change in China's MPI and examines the spillover effects. Indonesia exhibits reduced growth compared to the unconditional forecast due to the fixed stagnation in China's MPI, as China may have a significant trade or economic influence on Indonesia. Stabilization effects might dominate, leading to less variability compared to the unconditional forecast for Malaysia. Singapore may show a flatter trend, as its economy could be sensitive to changes in China. Effects might be dampened for Vietnam, reflecting a dependency on China's growth momentum.

Conditional forecasting allows small uncertainty (0.1%) in China's MPI values for the next 5 periods, reflecting more realistic constraints. This scenario introduces slight fluctuations in other countries' forecasts compared to the fixed scenario. The uncertainty reflects minor spillovers, adding variability to Indonesia, Malaysia, and Vietnam's MPI. Singapore, due to its high integration with the global economy, may show slightly larger forecast ranges.

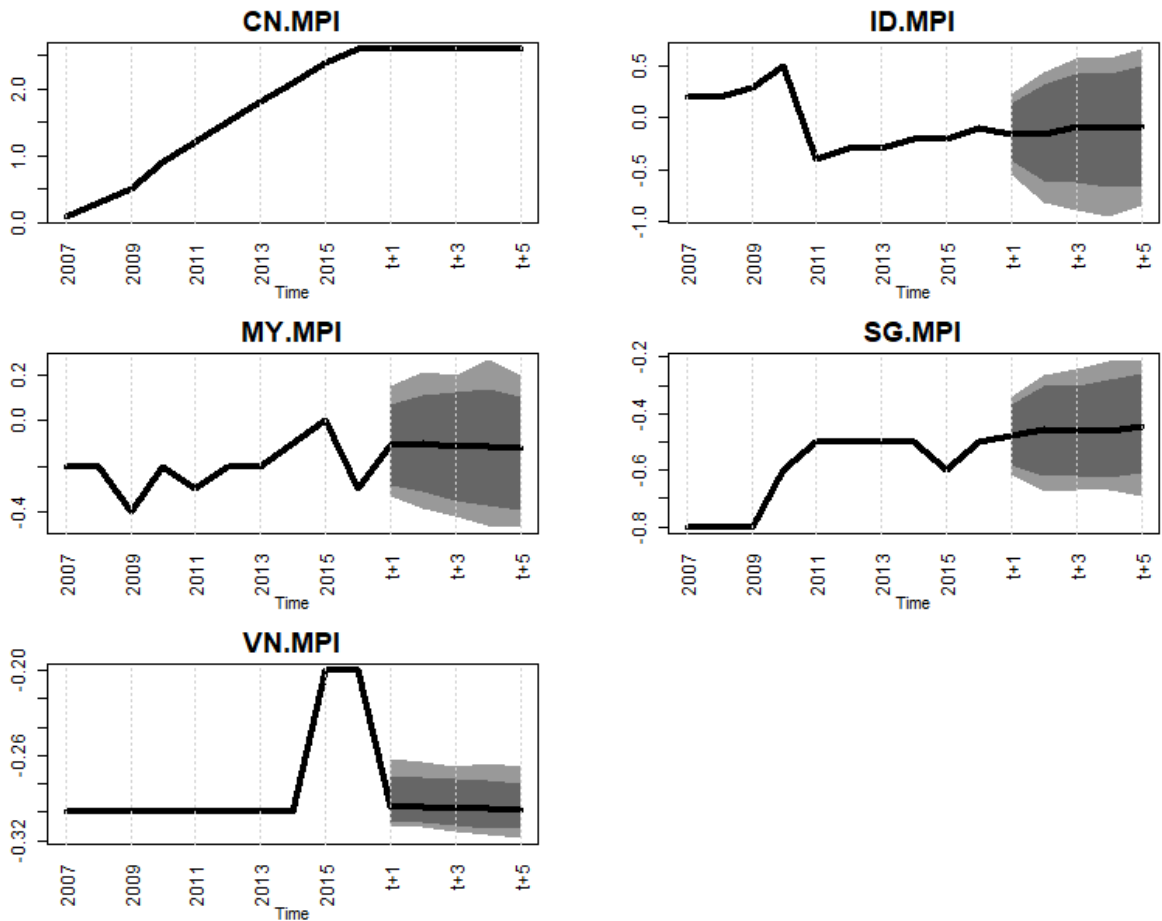

Figure 4. Conditional forecast with uncertain constraints

As seen in the unconditional forecast, China's MPI growth strongly influences regional peers, particularly Indonesia, Malaysia, and Vietnam. This reflects trade and economic linkages. Fixing CN.MPI removes variability, showing limited spillovers from China to others. Introducing uncertainty allows for more dynamic interactions, highlighting that even small shocks in China's MPI have ripple effects.

Table 5. Comparative Insights

| Feature | Unconditional | Conditional (Fixed) | Conditional (Uncertain) |
|---|---|---|---|
| China (CN.MPI) | Strong upward trend | No change (flat line) | Minimal fluctuation around the flat line |
| Uncertainty (All) | Widens over time | Reduced, especially for China's peers | Intermediate levels of uncertainty |
| Spillovers to Others | Larger effects due to China's growth | Dampened, reflecting China's stagnation | Minor effects from uncertainty in CN.MPI |
| MPI Variability | Higher for interconnected economies | Lower variability across countries | Moderate variability in response to uncertainty |

Figure 5 shows the impact of shocks on China's MPI and its effect on the MPI of other countries using generalized impulse responses.

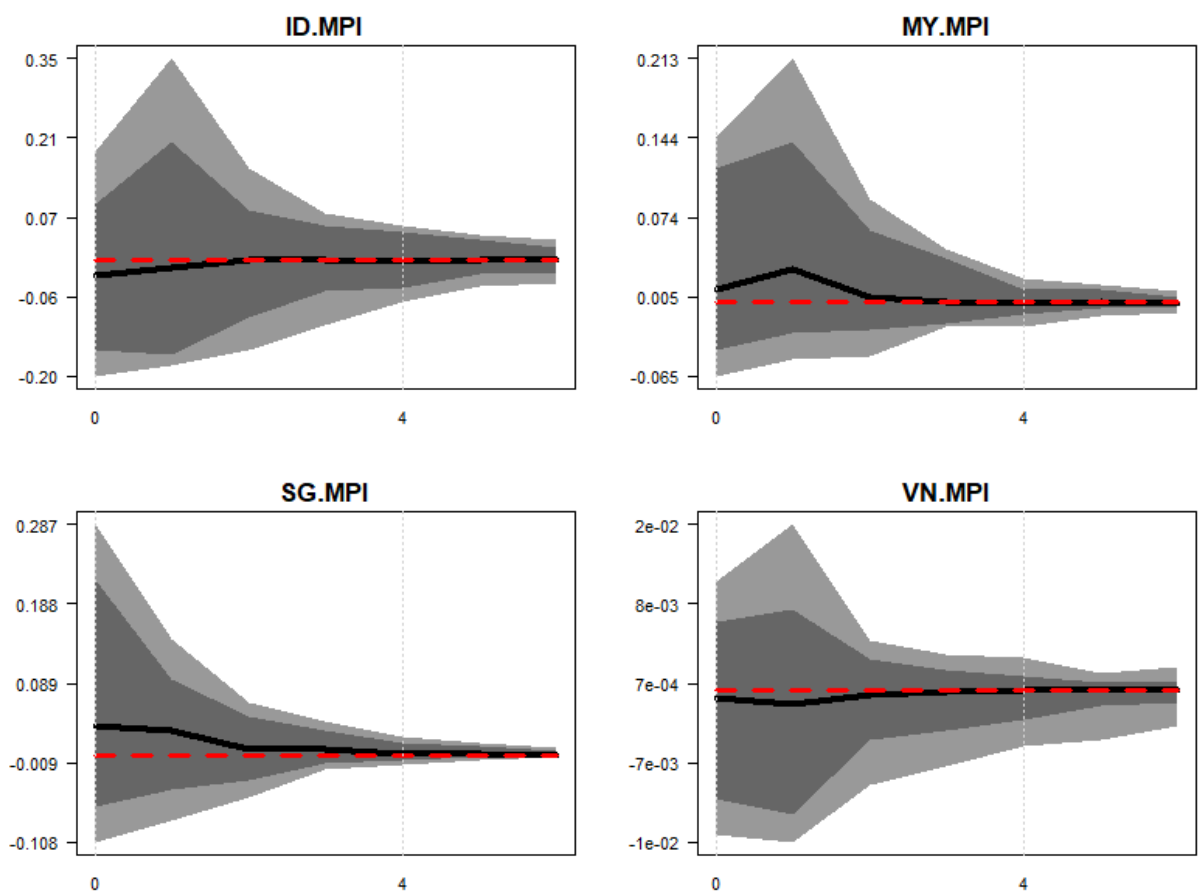

Figure 5. Generalized impulse responses for MPI

The figure demonstrates how shocks to China's MPI strongly influences Indonesia and Malaysia, reflecting significant spillovers likely linked to trade and production networks. Vietnam shows a pronounced dependency on China's MPI, underlining its sensitivity to China's economic policies. Impact of shocks on China's CPI, Exchange Rate, Exports, Imports, GDP per capita, Money Supply, and Unemployment and their effects on these variables for India, Malaysia, Singapore, and Vietnam are provided in the Appendix illustrated by generalized impulse response plots in Figure A1-A7.

## 5. Conclusions

The results obtained show several strong multi-way cointegration and causal relationships among the decision variables such as i) CPI with the changes in Exports and Imports for all countries, ii) Exchange Rate with Unemployment Rate in Malaysia, iii) Exports and Imports with Exchange Rate and Money Supply in China, iv) MPI with GDP per capita, Imports, and Unemployment in Singapore, and v) CPI with Imports and Exports, GDP per capita, and Money Supply in Vietnam.

Regarding the first research question addressing the spillover effects, the generalized impulse responses demonstrate how shocks to China propagate to other South Asian countries investigated here. For instance, a positive shock to China's MPI strongly influences Indonesia and Malaysia, reflecting significant spillovers linked to trade networks. When China's MPI is fixed or subjected to uncertainty, the forecast for the MPI of other countries change significantly, suggesting that China's policy variables have a strong influence on the other countries, illustrating the interconnectedness of economic policies across the region. Finally, the normalized trade matrix show that the spillovers are not uniform but depend on trade intensity, with Vietnam most affected by the changes in economy-wide policies of China.

Dynamic connectedness addressed by the second research question is also investigated in detail leading to the conclusion that the manufacturing hubs like China, Malaysia, and Vietnam exhibit interconnected dynamics that are likely driven by shared production chains and export dependencies. Dynamic connectedness is also evident in spillover mechanism shown in generalized impulse responses. Such as China's MPI significantly impacting Vietnam, Malaysia, and Indonesia, reflecting the cascading effects of shared manufacturing supply chains.

These findings are consistent with economic theory. Currency devaluations lead to inflation which forces governments to slow down the economy amid its negative effect on employment. Similarly, fluctuating levels of imports and exports alter the flows of foreign currency pushing policymakers to take subsequent steps to preserve the exchange rates around their real values.

The empirical results obtained provide several insights for policymakers as well as managers. First, statistical noise, which includes residuals and measurement and sampling mistakes, can make policy conclusions based only on the correlation of two variables misleading. This framework can be utilized by managers when making investment decisions to relocate their manufacturing facilities and policymakers when formulating economic policies to promote international trade. Policymakers in Malaysia, Vietnam, and Indonesia should closely monitor economic policies of China, as spillovers from China's shocks can significantly affect their economies. Conditional forecasts demonstrate that stabilizing policy variables of China can reduce uncertainty in regional economic outcomes. Regional collaboration is essential to manage shared risks and leverage opportunities within the manufacturing ecosystem. Singapore's role as a trade hub can be further strengthened to mitigate risks associated with heavy reliance on China.

There are inherent constraints in our research due to the data availability across countries. One such example is the Purchasing Managers' Index (PMI), which indicates the direction of economic developments in the manufacturing and service sectors. It can be expanded to include the other nations that are being studied. Investigating the mechanisms driving stronger spillovers such as trade intensity and supply chain linkages can help refine policy strategies. Examining how specific policy shocks propagate through these manufacturing hubs could also provide actionable insights for regional cooperation.

**Compliance with Ethical Standards**

This article does not contain any studies with human participants performed by the authors.

**Data Availability Statement**

The data supporting the findings of this study are available at the Asian Development Bank Key Indicators Database (https://kidb.adb.org/).

**Competing Interests**

We wish to confirm that there are no known conflicts of interest associated with this publication and there has been no significant financial support for this work that could have influenced its outcome.

**Funding**

The authors declare that no funds, grants, or other support were received during the preparation of this manuscript.

**Appendix**

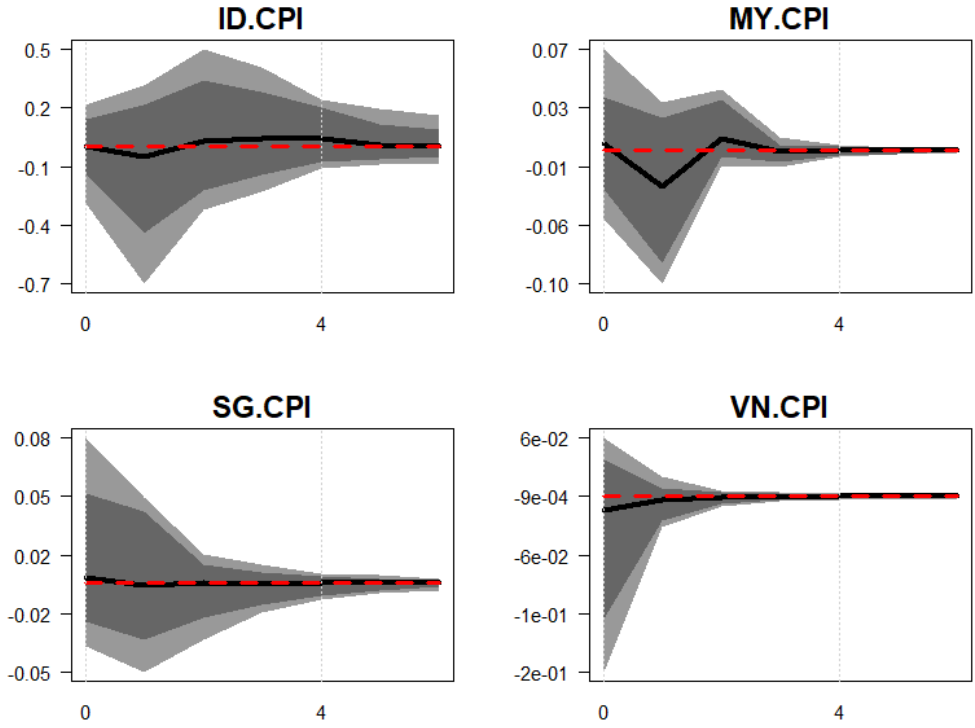

Figure A1. Generalized impulse responses for CPI

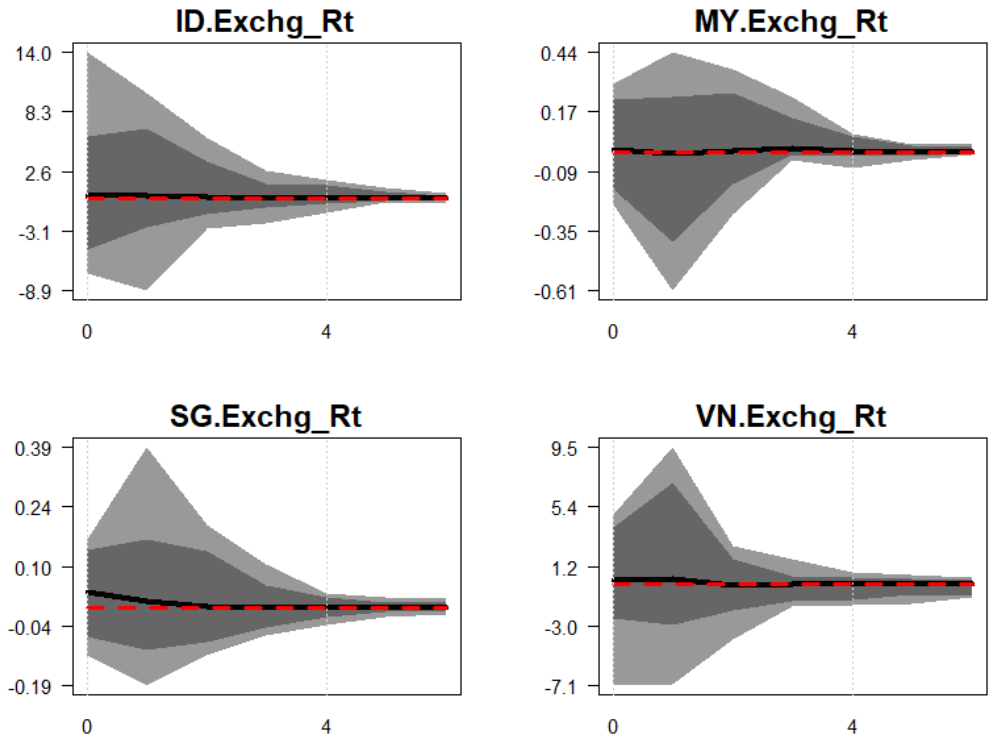

Figure A2. Generalized impulse responses for Exchange Rate

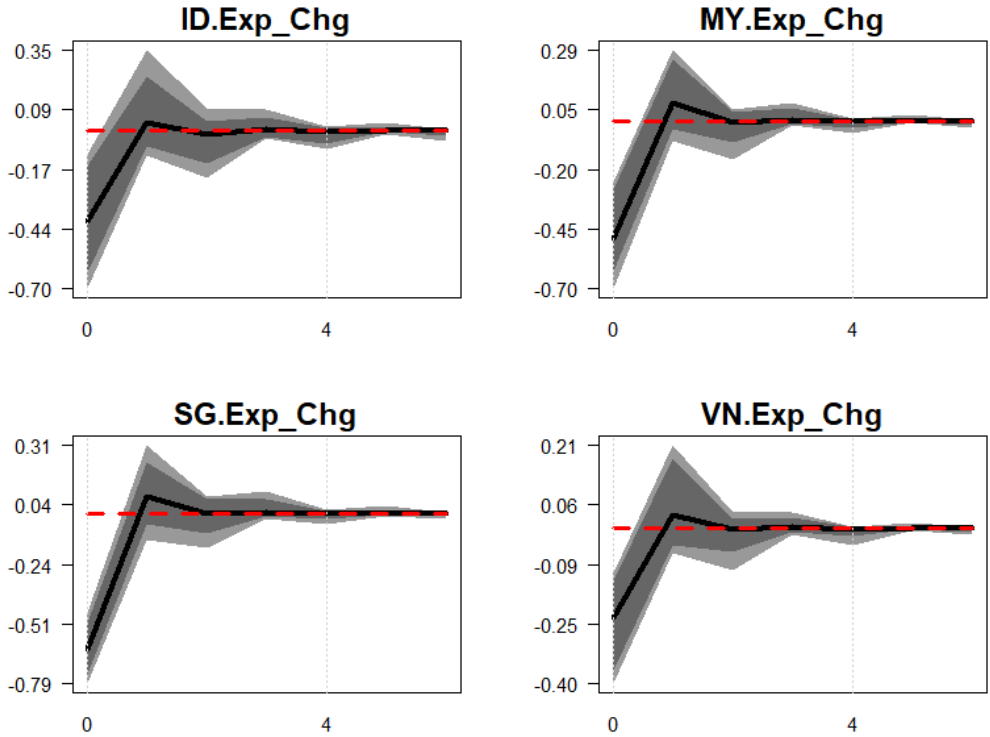

Figure A3. Generalized impulse responses for Exports

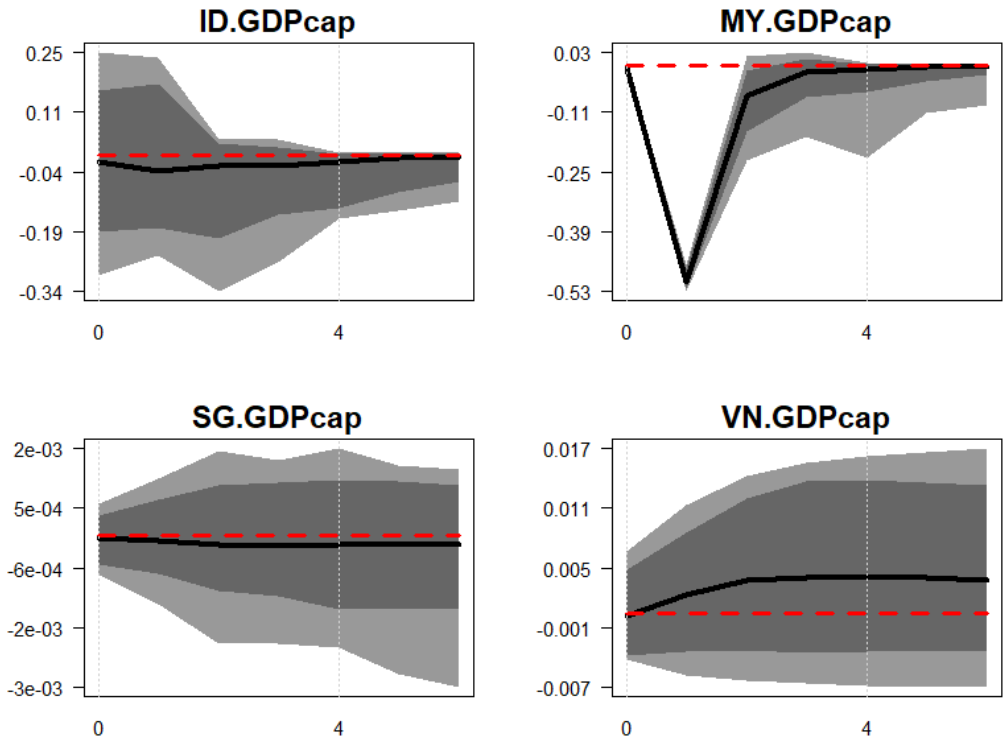

Figure A4. Generalized impulse responses for GDP per capita

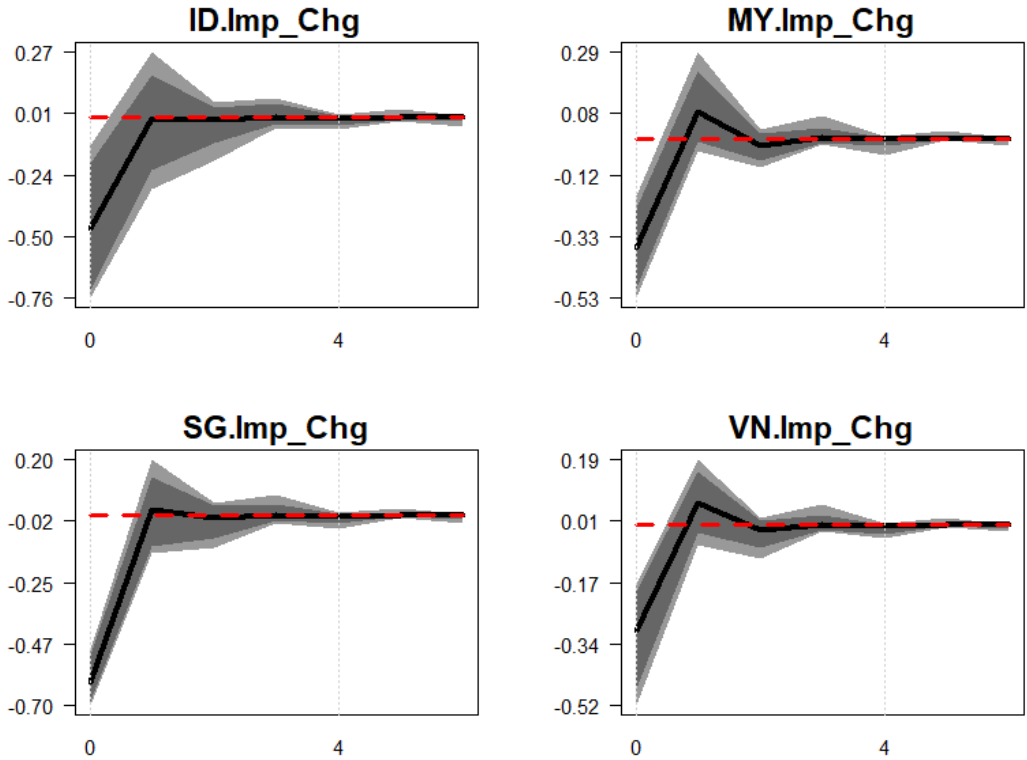

Figure A5. Generalized impulse responses for Imports

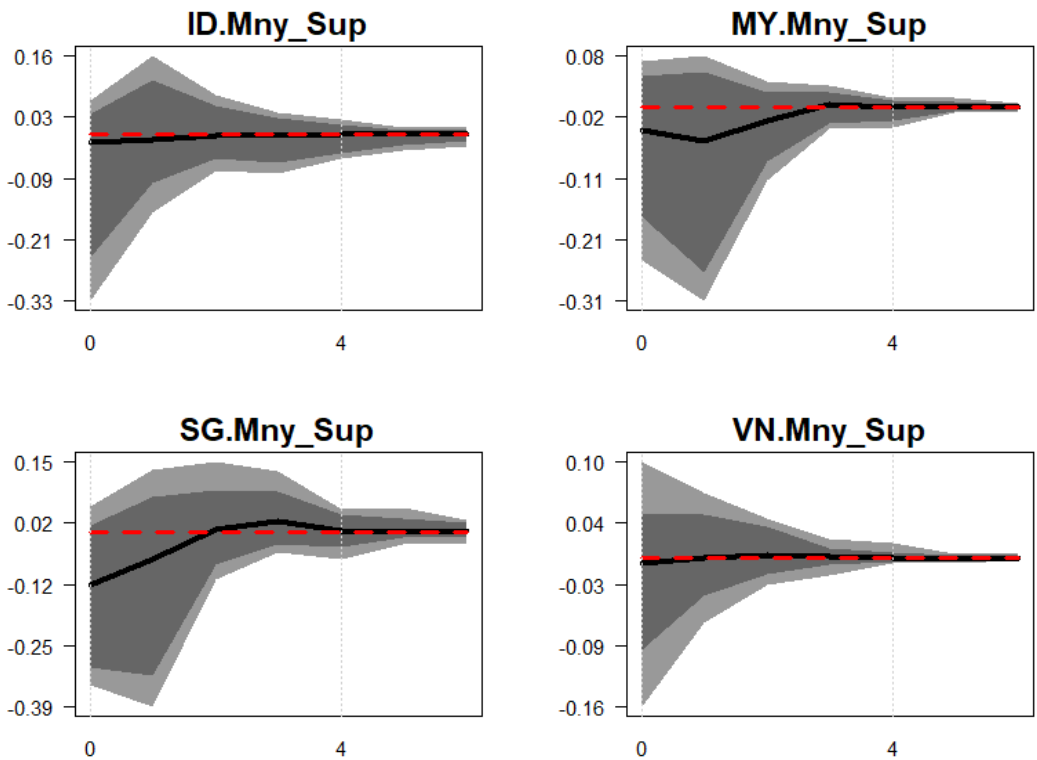

Figure A6. Generalized impulse responses for Money Supply

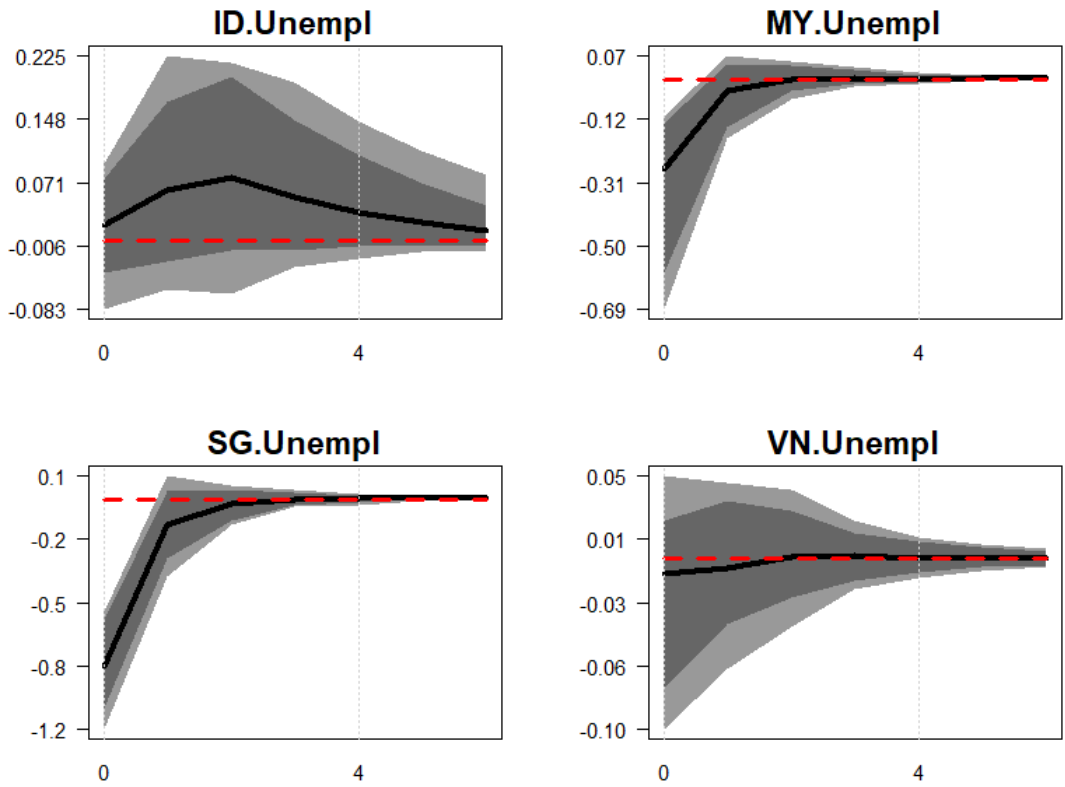

Figure A7. Generalized impulse responses for Unemployment